\documentclass[sigconf]{acmart}

\usepackage{enumitem}
\usepackage{bm}
\usepackage{xspace}
\usepackage{color}
\usepackage{mathtools}
\usepackage{subcaption}
\usepackage{booktabs}
\usepackage{multirow}
\usepackage{adjustbox}
\usepackage[ruled, vlined]{algorithm2e}
\usepackage[font=small,labelfont=bf,tableposition=top]{caption}
\usepackage{tikz}
\usepackage{balance}

\setlist[itemize]{leftmargin=18pt}
\setlist[enumerate]{leftmargin=24pt}
\newlength{\Oldarrayrulewidth}

\newcommand{\vpara}[1]{\vspace{0.08in}\noindent\textbf{#1}}
\newcommand{\hide}[1]{} 

\newcommand\hmm[1]{\ifnum\ifhmode\spacefactor\else2000\fi>1000 \uppercase{#1}\else#1\fi}

\newcommand{\ie}{{\sl i.e.\xspace}}
\newcommand{\etc}{{\sl etc.}}

\newcommand{\trans}{^\top}

\newcommand{\mc}{\mathcal}

\renewcommand{\phi}{\varphi}
\newcommand{\mbf}{\mathbf{f}}
\newcommand{\mbg}{\mathbf{g}}
\newcommand{\bmmu}{\bm{\mu}}

\newcommand{\heer}{\textsc{HEER}\xspace}
\newcommand{\stan}{\textit{Stan}\xspace}
\newcommand{\anglee}{\textit{Ang Lee}\xspace}
\newcommand{\musical}{\textit{musical}\xspace}

\begin{document}
\title{Easing Embedding Learning by Comprehensive Transcription of Heterogeneous Information Networks}

\author{
Yu Shi\footnotemark[1]\ \ \ \ 
Qi Zhu\footnotemark[1]\ \ \ \ 
Fang Guo\ \ \ \ 
Chao Zhang\ \ \ \ 
Jiawei Han\\
{\fontsize{10pt}{12pt}\selectfont{\text{University of Illinois at Urbana-Champaign, Urbana, IL USA}}}   \\
{\fontsize{10pt}{12pt}\selectfont{\text{\{yushi2, qiz3, fangguo1, czhang82, hanj\}@illinois.edu}}}\\
}


\begin{abstract}
Heterogeneous information networks (HINs) are ubiquitous in real-world applications.
In the meantime, network embedding has emerged as a convenient tool to mine and learn from networked data.
As a result, it is of interest to develop HIN embedding methods.
However, the heterogeneity in HINs introduces not only rich information but also potentially incompatible semantics, which poses special challenges to embedding learning in HINs.
With the intention to preserve the rich yet potentially incompatible information in HIN embedding, we propose to study the problem of comprehensive transcription of heterogeneous information networks.
The comprehensive transcription of HINs also provides an easy-to-use approach to unleash the power of HINs, since it requires no additional supervision, expertise, or feature engineering.
To cope with the challenges in the comprehensive transcription of HINs, we propose the \heer algorithm, which embeds HINs via edge representations that are further coupled with properly-learned heterogeneous metrics.
To corroborate the efficacy of \heer, we conducted experiments on two large-scale real-words datasets with an edge reconstruction task and multiple case studies. 
Experiment results demonstrate the effectiveness of the proposed \heer model and the utility of edge representations and heterogeneous metrics.
The code and data are available at \url{https://github.com/GentleZhu/HEER}.
\end{abstract}

\copyrightyear{2018} 
\acmYear{2018} 
\setcopyright{acmlicensed}

\acmConference[KDD '18]{The 24th ACM SIGKDD International Conference on Knowledge Discovery & Data Mining}{August 19--23, 2018}{August 19--23, 2018, London, United Kingdom}
\acmPrice{15.00}
\acmDOI{http://dx.doi.org/10.1145/3219819.3220006}
\acmISBN{978-1-4503-5552-0/18/08}


%
%




\maketitle

\vspace{-9pt}
\begin{small}
\textbf{ACM Reference format:}\\
Yu Shi, Qi Zhu, Fang Guo, Chao Zhang, and Jiawei Han. 2018. Easing Embedding Learning by Comprehensive Transcription of Heterogeneous Information Networks. In \textit{Proceedings of the 24th ACM SIGKDD International Conference on Knowledge Discovery and Data Mining, London, United Kingdom, August 19--23, 2018 (KDD '18)}. DOI: 10.1145/3219819.3220006
\end{small}
\vspace{-3pt}

{
\renewcommand{\thefootnote}{\fnsymbol{footnote}}
\footnotetext[1]{These authors contributed equally to this work.}
}


\section{Introduction}\label{sec::introduction}
\begin{figure}[t]
 \centering\includegraphics[width=0.9\linewidth]{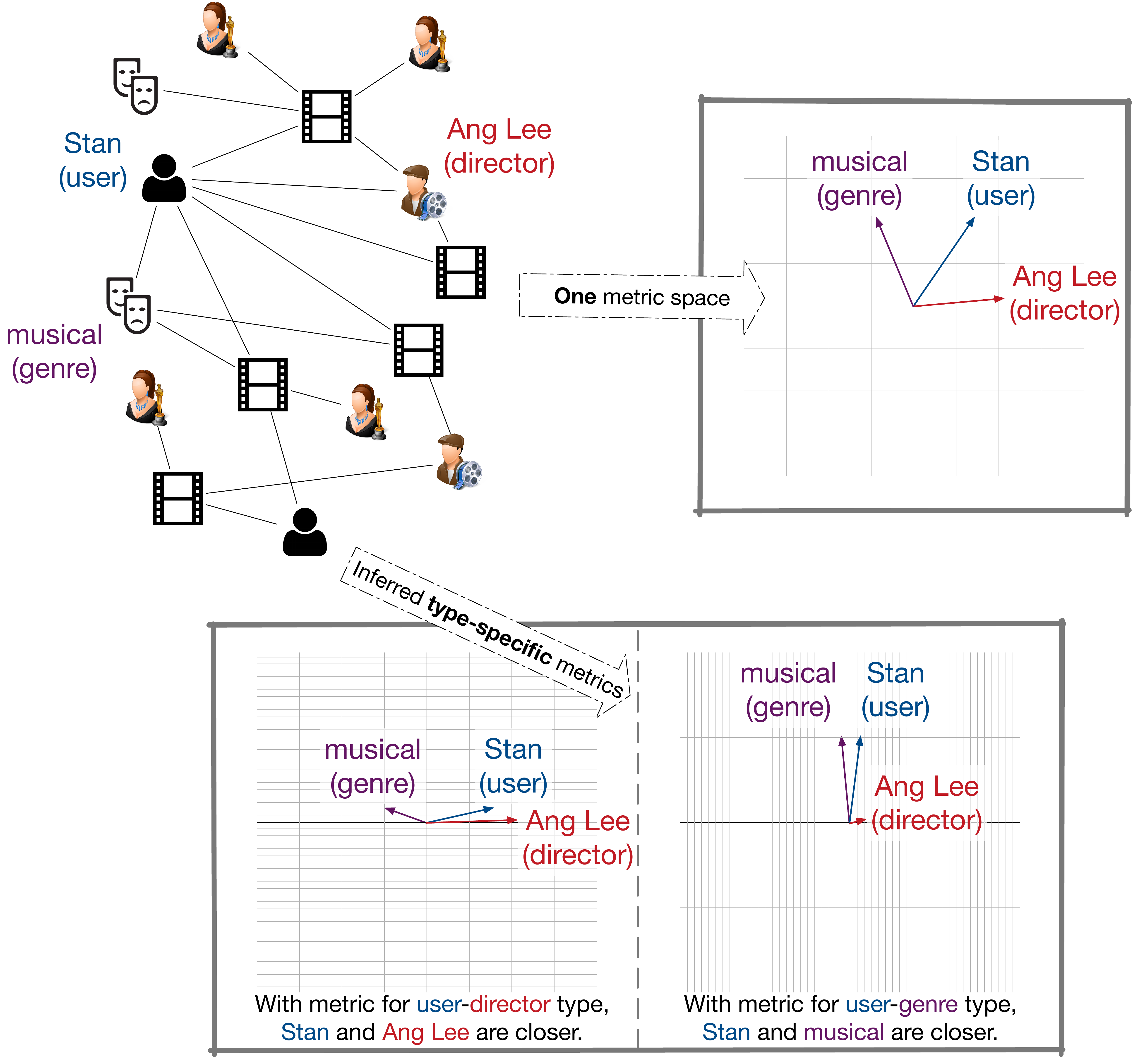}
 \caption[]{
To preserve the rich information in HIN embedding, properly handling the incompatibility introduced by the heterogeneity is necessary.
The upper left part of the figure gives a toy movie-reviewing HIN, where users review movies and list certain directors, actors, genres as their favorites. 
\stan likes both \musical and movies directed by \anglee. 
If all nodes were embedded to one metric space, \stan would be close to neither \musical nor \anglee due to the dissimilarity between \musical and \anglee.  
This results in information loss in the embedding learning process.
However, we can alleviate this problem by employing edge representation and inferring edge-type--specific metrics, so that \stan can be close to both \musical and \anglee under their respective metrics, while not necessarily dragging \musical and \anglee closer.
The two metrics shown in the lower figure can be achieved by linearly transforming the metric space in the upper right figure.
 }\label{fig::intuition}
\end{figure}

Heterogeneous information networks (HINs) have received increasing attention in the past decade due to its ubiquity and capability of representing rich information~\cite{shi2017survey, sun2013mining}.
Meanwhile, network embedding has emerged as a scalable representation learning method~\cite{dong2017metapath2vec, grover2016node2vec, perozzi2014deepwalk, ribeiro2017struc2vec, tang2015pte, tang2015line, wang2016structural}.
Network embedding learns low-dimensional vector representations for nodes to encode their semantic information in the original network. 
The vectorized representations can be easily combined with off-the-shelf machine learning algorithms for various tasks such as classification and link prediction~\cite{perozzi2014deepwalk, grover2016node2vec, tang2015line, hamilton2017representation}, which provides a convenient approach for researchers and engineers to mine and learn from the networked data. 
To marry the advantages of HINs and network embedding, researchers have recently started to explore methods to embed heterogeneous information networks~\cite{shang2016meta, dong2017metapath2vec, fu2017hin2vec, chang2015heterogeneous, gui2016large, tang2015pte, shi2018aspem}, and have demonstrated the effectiveness of HIN embedding in applications including author identification~\cite{chen2017task}, name disambiguation~\cite{zhang2017name}, proximity search~\cite{liu2017semantic}, event detection~\cite{zhang2017triovecevent}, \etc

However, the heterogeneity in HINs brings in not only rich information but also potentially incompatible semantics, which poses special challenges to embed heterogeneous information networks.
Take the movie-reviewing network in Figure~\ref{fig::intuition} as an example, where users review movies and list certain actors, directors, and genres as their favorites.
Suppose user \stan likes both movies directed by \anglee{} {(director)} and \musical{} {(genre)}.
Since \anglee has never directed any \musical, nor is he semantically similar to \musical, if this HIN were embedded into \textit{one} metric space, \musical and \anglee would be distant from each other, while the user \stan would not be simultaneously close to both of them, due to the triangle inequality property of metric spaces.
We have also observed different extents of such incompatibility from real-world data as to be discussed in Section~\ref{sec::observation}, which is consistent with the observation that different extents of correlation can exist within one HIN as per existing study~\cite{shi2017prep}.
As a result, it can be expected that an algorithm would generate better embeddings if it additionally models such semantic incompatibility.
We hence study the problem of \textit{comprehensive transcription of heterogeneous information networks}, which purely aims to transcribe the rich and potentially incompatible information from HINs to the embeddings, without involving additional expertise, feature engineering, or installation of supervision.

With HINs comprehensives transcribed, one can again pipe the \textit{unsupervisedly} learned embeddings to off-the-shelf machine learning algorithms for a wide range of applications.
Therefore, beyond the capability of preserving rich information, another motivation to study comprehensive transcription of HINs is to provide an easy-to-use approach to unleash the power of HINs in a wide variety of applications with no expertise or supervision required in the embedding learning process.

Traditional homogeneous network embedding methods~\cite{grover2016node2vec, perozzi2014deepwalk, ribeiro2017struc2vec, tang2015line, wang2016structural} treat all the nodes and edges equally regardless of their types, which do not capture the essential heterogeneity of HINs.
A couple of methods have recently been studied for embedding heterogeneous information networks~\cite{shang2016meta, dong2017metapath2vec, fu2017hin2vec, chang2015heterogeneous, gui2016large, tang2015pte, shi2018aspem}.
Many of them build their algorithms on top of a set of meta-paths~\cite{shang2016meta, dong2017metapath2vec}, which often require users to specify the meta-paths or leverage supervision to make the meta-path selection.
However, a set of meta-paths specified or selected in this way often only reflects certain aspects of the HIN or is suitable for specific tasks.
As a result, they are not always capable of transcribing HINs comprehensively.
These methods are not as easy-to-use either because it involves the additional meta-path generation process that entails expertise or supervision.
Besides using meta-paths, some approaches have been proposed to embed specific kinds of HINs~\cite{gui2016large, tang2015pte} for certain tasks or HINs with additional side information~\cite{chang2015heterogeneous}.
These methods cannot be applied to comprehensively transcribe general HINs.
Additionally, most existing HIN embedding methods~\cite{shang2016meta, dong2017metapath2vec, gui2016large, tang2015pte} employ only one metric space for embedding learning.
This approach may suit downstream tasks that are related to certain partial information of an HIN with compatible semantics but could lead to information loss if the objective is to comprehensively transcript the entire HIN.

The problem of comprehensive transcription of HINs is challenging because it requires the modeling of heterogeneity that can be complex and incompatible.
Besides, without the availability of supervision, proposed solutions need to capture the latent structure of the HINs and distinguish potentially incompatible semantics in an unsupervised way.
To cope with these challenges, we propose \textbf{h}eterogeneous information network \textbf{e}mbedding via \textbf{e}dge \textbf{r}epresentations, which is henceforth referred to as \heer.
\heer builds edge embeddings atop node embeddings, which are further coupled with inferred heterogeneous metrics for each edge type.
The inferred metrics capture which dimensions of the edge embeddings are more important for the semantic carried by their corresponding edge types.
In turn, the information carried by edges of different types updates the node embeddings and edge embeddings with emphases on different type-specific manifolds.
In this way, we can preserve different semantics even in the presence of incompatibility.
Still take the movie-reviewing network as example, by adopting heterogeneous metrics as in the lower part of Figure~\ref{fig::intuition}, \stan could be close to both \musical {(genre)} and \anglee {(director)} under their respective metrics.
Furthermore, the heterogeneous metrics are inferred by fitting the input HIN, so that semantic incompatibility is captured without additional supervision.

Specifically, with the availability of edge representations and coupled metrics, we derive loss function that reflects both the existence and the type of an edge. 
By minimizing the loss, the node embeddings, edge embeddings, and heterogeneous metrics are updated simultaneously, and thereby retain the heterogeneity in the input HIN.
Different extents of incompatibility can also be modeled, where the more compatible two edge types are, the more similar their corresponding metrics would be.

Lastly, we summarize our contributions as follows:
\begin{enumerate}
\item 
We propose to study the problem of comprehensive transcription of HINs in embedding learning, which preserves the rich information in HINs and provides an easy-to-use approach to unleash the power of HINs.
\item
We identify that different extents of semantic incompatibility exist in real-world HINs, which pose challenges to the comprehensive transcription of HINs.
\item
We propose an algorithm, \heer, for the comprehensive transcription of HINs that leverages edge representations and heterogeneous metrics.
\item
Experiments with real-world large-scale datasets demonstrate the effectiveness of \heer and the utility of edge representations and heterogeneous metrics.
\end{enumerate}


\section{Related Work}\label{sec::related-work}
\vpara{Homogeneous network embedding.}
Meanwhile, network embedding has emerged as an efficient and effective representation learning approach for networked data~\cite{grover2016node2vec, ou2016asymmetric, perozzi2014deepwalk, ribeiro2017struc2vec, tang2015line, wang2016structural, hamilton2017representation, dai2016discriminative,zhang2017weisfeiler,ribeiro2017struc2vec}, which significantly spares the labor and sources in transforming networks into features that are more machine-actionable.
Early network embedding algorithms start from handling the simple, homogeneous networks, and many of them trace to the skip-gram model~\cite{mikolov2013distributed} that aims to learn word representations where words with similar context have similar representation~\cite{grover2016node2vec, perozzi2014deepwalk, ribeiro2017struc2vec, tang2015line}. 
Besides skip-gram, algorithms for preserving certain other homogeneous network properties have also been studied~\cite{ou2016asymmetric, wang2016structural, xu2018exploring, niepert2016learning, velivckovic2017graph, kipf2016semi}.
The use of edge representations for homogeneous network embedding is discussed in a recent work~\cite{abu2017learning}, but such edge representations are designed to distinguish the direction of an edge, instead of encoding richer semantics such as edge type in our case.

\vpara{Heterogeneous network embedding.}
Heterogeneous information network (HIN) has been extensively studied since the past decade for its ubiquity in real-world data and efficacy in fulfilling tasks, such as classification, clustering, recommendation, and outlier detection~\cite{shi2017survey, sun2013mining, sun2009ranking, yu2014personalized, zhuang2014mining}. 
To marry the advantages of HIN and network embedding, a couple of algorithms have been proposed very recently for embedding learning in heterogeneous information networks~\cite{shang2016meta, dong2017metapath2vec, fu2017hin2vec, chang2015heterogeneous, gui2016large, tang2015pte, shi2018aspem}.
One line of work first uses human expertise or supervision to select meta-paths for a given task or limit the scope of candidate meta-paths, and then proposes methods to transfer the semantics encoded in meta-paths to the learned embedding~\cite{shang2016meta, dong2017metapath2vec, fu2017hin2vec}.
While this direction has been showed to be effective in solving problems that fit the semantics of the chosen meta-paths, it differs from the research scope of ours because they mostly focus on providing quality representations for downstream tasks concerning the node types on the two ends of chosen meta-paths, while we aim at developing methods to transcribe the entire HIN to embeddings as comprehensively as possible.
Beyond meta-paths, some approaches have been proposed to embed specific kinds of HINs~\cite{gui2016large, tang2015pte} with specific objectives such as representing event data or learning predictive text embeddings. 
Some other approaches study HINs with additional side information~\cite{chang2015heterogeneous} that cannot be generalized to all HINs.
Besides, all of these approaches embed the input HIN into only one metric space.
Embedding in the context of HIN has also been studied for tasks with additional supervision~\cite{chen2017task,  liu2017semantic, pan2016tri}. 
These methods either yield features specific to given tasks, and are outside of the scope of unsupervised HIN embedding that we study.

A recent study~\cite{shi2018aspem} proposes a method by decomposing an HIN into multiple aspects before learning embedding, which also attains quality representations of HINs by alleviating the information loss arising from the rich, yet heterogeneous, and potentially conflicting semantics within the given networks.
However, this approach embeds the derived aspects independently and completely forbids joint learning across aspects while our proposed method allows network components of varied compatibility to collaborate to different extents in the joint learning process.


\section{Preliminaries}\label{sec::prob-def}
In this section, we define related concepts and notations. 

\begin{definition}[Heterogeneous Information Network]
An \textbf{information network} is a directed graph $G = (\mc{V}, \mc{E})$ with a node type mapping $\phi: \mc{V} \rightarrow \mc{T}$ and an edge type mapping $\psi: \mc{E} \rightarrow \mc{R}$. 
Particularly, when the number of node types $|\mc{T}| > 1$ or the number of edge types $|\mc{R}| > 1$, the network is called a \textbf{heterogeneous information network} (HIN).
\end{definition}

\begin{figure}[t]
 \centering\includegraphics[width=.8\linewidth]{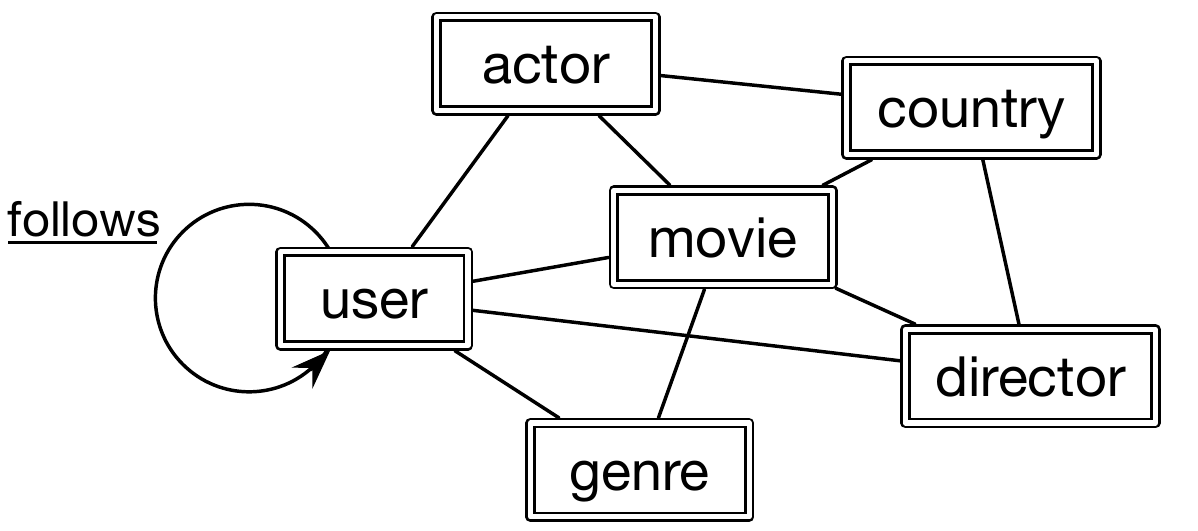}
 \caption{The schema of a toy movie-reviewing HIN with six node types, seven undirected edge types, and one directed edge type.}\label{fig::toy-schema}
\end{figure}

Given the typed essence of HINs, the network schema $\tilde{G} = (\mc{T}, \mc{R})$~\cite{sun2013mining} is used to abstract the meta-information regarding the node types and edge types in an HIN. 
Figure~\ref{fig::toy-schema} illustrates the schema of a toy movie-reviewing HIN.

In addition, we require that only one node type can be associated with a certain end of an edge type. 
That is, once an edge type is given, we would deterministically know the node types on its two ends. 
As an example, consider two edges with one representing director \textit{Fatih Akin} living in \textit{Germany} and another representing movie \textit{In the Fade} being produced in \textit{Germany}. 
Such requirement implies that these two edges must have distinct types -- \textit{livesIn} and \textit{isProducedIn} -- instead of just one type -- \textit{isIn}.
For edge type $r \in \mc{R}$, we denote $\mc{P}^r \coloneqq \big\{(u, v) \in \mc{V} \times \mc{V} \,\big|\, \phi(u) \!  \sim \! r \! \sim \! \phi(v)) \big\}$, where $\phi(u) \!  \sim \! r \! \sim \! \phi(v)$ means the node type pair $(\phi(u), \phi(v)))$ is consistent with edge type $r$. Additionally, define $\mc{P}^r_{u \ast} \coloneqq \big\{\tilde{v} \in \mc{V} \big| (u, \tilde{v}) \in \mc{P}^r \big\}$ and $\mc{P}^r_{\ast v} \coloneqq \big\{\tilde{u} \in \mc{V} \big| (\tilde{u}, v) \in \mc{P}^r \big\}$.

Moreover, when the network is weighted and directed, we use $W^{(r)}_{uv}$ to denote the weight of an edge $e \in \mc{E}$ with type $r \in \mc{R}$ that goes out from node $u$ toward $v$.
$D^{O(r)}_u$  and $D^{I(r)}_u$ respectively represent the outward degree of node $u$ (\ie, the sum of weights of all type-$r$ edges going outward from $u$) and the inward degree of node $u$ (\ie,  the sum of weights of all type-$r$ edges going inward to $u$). 
For an unweighted edge, $W^{(r)}_{uv}$ is trivially $1$. 
For an undirected edge, we always have $W^{(r)}_{uv} = W^{(r)}_{vu}$ and $D^{O(r)}_u = D^{I(r)}_u$.

\begin{definition}[Node and Edge Representations in HIN Embedding] 
Given an HIN $G = (\mc{V}, \mc{E}; \phi, \psi)$, the problem of HIN embedding via edge representations learns a node embedding mapping $f: \mc{V} \rightarrow \mathbb{R}^{d_\mc{V}}$ and an edge embedding mapping $f: \mc{V} \times \mc{V} \rightarrow \mathbb{R}^{d_\mc{E}}$, where $d_\mc{V}$ and $d_\mc{E}$ are the dimensions for node and edge embeddings, respectively. A node $u \in \mc{V}$ is thereby represented by a node embedding $\mbf_u \coloneqq f(u)$ and a node pair $(u, v) \in \mc{V} \times \mc{V}$ is represented by an edge embedding $\mbg_{uv} \coloneqq g(u, v)$.
\end{definition}

With this definition, a node pair has its edge embedding even if no edge of any type has been observed between them. 
On the other hand, it is possible for node pair $(u, v)$ to be associated by multiple edges with different types, and we expect edge embedding $\mbg_{uv}$ to encapsulate such information of an HIN.

Finally, we define the problem of comprehensive transcription of a heterogeneous information network in embedding learning.
\begin{definition}[Comprehensive Transcription of an HIN]
The comprehensive transcription of an HIN aims to learn the representations of the input HIN that retains the rich in the HIN as comprehensively as possible, in an approach that does not require additional expertise, feature engineering, or supervision.
\end{definition}


\section{Varied Extents of Incompatibility due to Heterogeneity}\label{sec::observation}
In this section, we look into the incompatibility in HINs using real-world data, and we take DBLP as an example.

DBLP is a bibliographical information network in the computer science domain~\cite{tang2008arnetminer}, where authors write papers that are further associated with nodes of other attribute types.
Since the measurable incompatibility in an HIN arises from the co-existence of multiple edge types, we dive down to the minimal case that involves two different edge types ($r_1$ and $r_2$) joined by a common node type ($t$).
To quantify the incompatibility for this minimal case, we use the widely used generalized Jaccard coefficient to measure the similarity between the node groups reachable from a given node of type $t$ via the two edge types.
Specifically, given node $u$ of type $t$, the Jaccard coefficient for edge types $r_1$ and $r_2$ is given by $J(u; r_1, r_2) \coloneqq \frac{\min_{\phi(v) = t} \{l(u, v; r_1), l(u, v; r_2)\}}{\max_{\phi(v) = t} \{l(u, v; r_1), l(u, v; r_2)\}}$, where $l(u, v; r) \coloneqq \bm{P}^r_{u,:}(\bm{P}^r_{v,:})\trans$ is the reachability between nodes $u$ and $v$ via edge type $r$ and $\bm{P}^r$ is the row-normalized adjacency matrix of edge type $r$.
Generalized Jaccard coefficient has a range of $[0, 1]$, and greater value implies more similarity, or equivalently, less incompatibility.

\begin{figure}[t]
  \centering
  \begin{subfigure}[t]{0.58\linewidth}
    \centering\includegraphics[width=\linewidth]{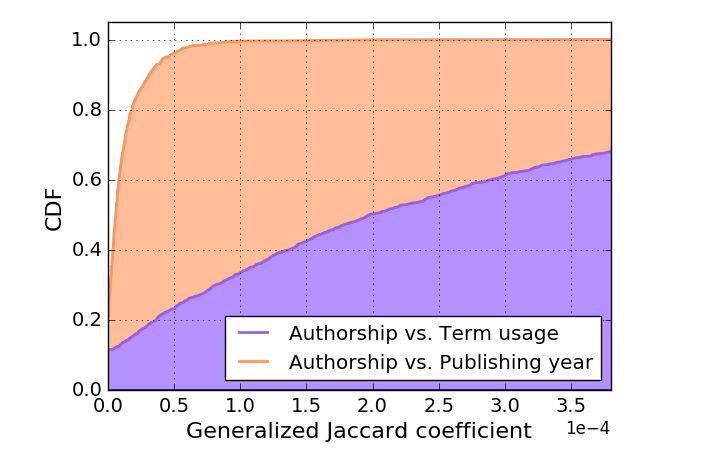}
    \caption{The CDF of the generalized jaccard coefficients for two pairs of edge types.}\label{fig::jaccard}
  \end{subfigure}
  \hspace{2pt}
  \begin{subfigure}[t]{0.38\linewidth}
    \centering\includegraphics[width=\linewidth]{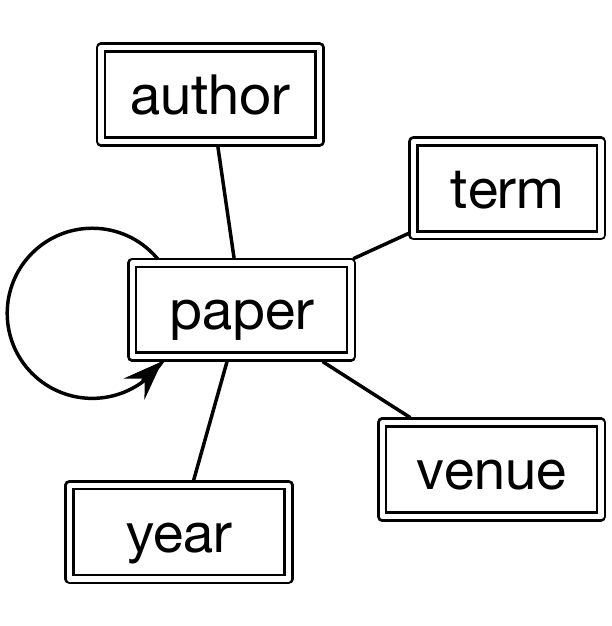}
    \caption{The schema of the DBLP network.}\label{fig::dblp-schema}
  \end{subfigure}
  \caption{Varied extents of incompatibility exist between different pairs of edge types in the DBLP network.}
\end{figure}

As an example, we consider four node types -- author, paper, key term, and year -- and two pairs of edge types -- (i) authorship vs. publishing year of papers and (ii) authorship vs. term usage of papers.
We illustrate the distributions over Jaccard coefficient using cumulative distribution function (CDF) for each of the two pairs in Figure~\ref{fig::jaccard}.
It can be seen that over $95\%$ of nodes have a generalized Jaccard coefficient smaller than $5e^{-5}$ between authorship and publishing year, while less than $25\%$ of nodes fall in the same category when it comes to authorship vs. term usage.
In other words, we observe more incompatibility between authorship and publishing year than between authorship and term usage.
However, this relationship is actually not surprising because papers published in the same year can be authored by any researchers who are active at that time, while key terms associated to certain research topics are usually used by authors focusing on these topics.
With the presence of such varied extent of incompatibility, we would expect an embedding algorithm tailored for comprehensive transcription of HINs to be able to capture this semantic subtlety in HINs.

\begin{figure*}[t]
 \centering\includegraphics[width=\linewidth]{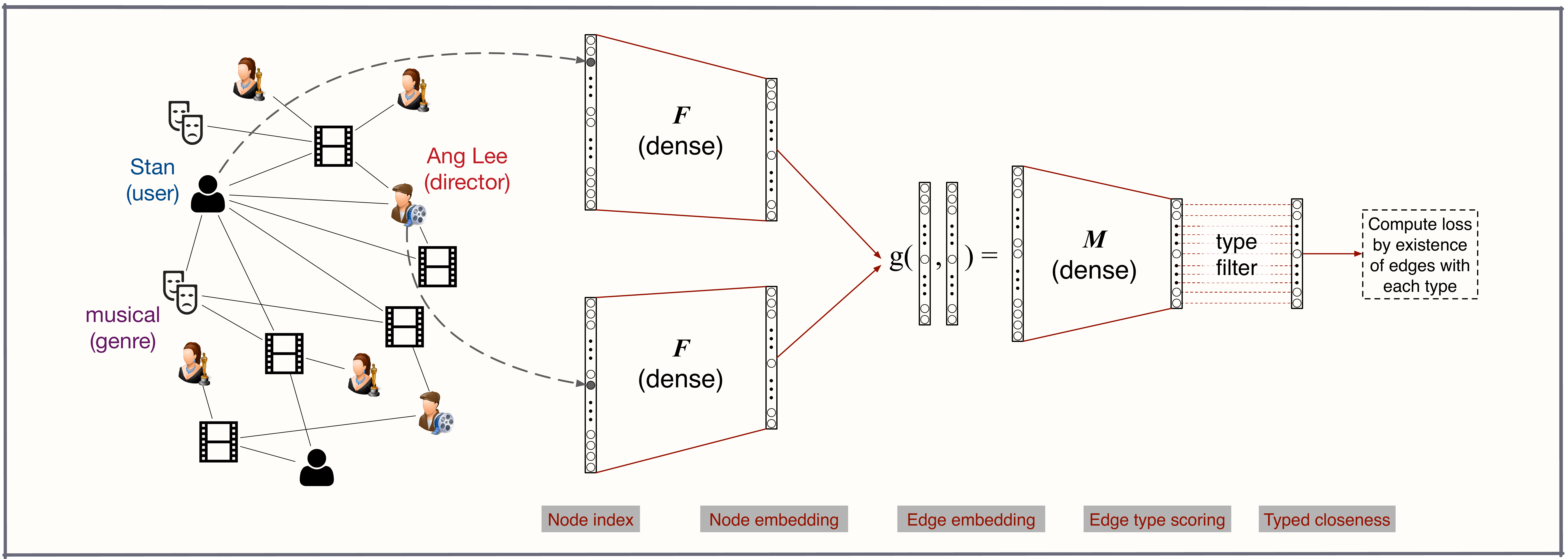}
 \caption[]{An illustration of the \heer architecture for learning HIN embedding via edge representation.}\label{fig::nn}
\end{figure*}

In fact, by employing edge representation and heterogeneous metrics, the inferred metrics could be learned to be different for incompatible edge types.
In turn, the information carried by these two edge types would be updating the node embeddings and edge embeddings with emphases on different manifolds.
On the other hand, the subtlety of the different extent of incompatibility could also be captured in a way that the more compatible two edge types are, the more similar their inferred metrics should be.



\section{Proposed Method}\label{sec::modeling}
To provide an general-purpose, easy-to-use solution to HIN embedding, we describe the \heer model in this section, where \heer stands for \textbf{H}eterogeneous Information Network \textbf{E}mbedding via \textbf{E}dge \textbf{R}epresentations. Afterward, the model inference method is described subsequently.


\subsection{The \heer Model}
A learned embedding that effectively encodes the semantics of an HIN should be able to reconstruct this HIN.
With the use of edge representation, we expect the embedding to infer not only the existence but also the type of edge between each pair of nodes.
For edge type $r \in \mc{R}$, we formulate the \textbf{typed closeness} of node pair $(u, v)$ atop their edge embedding $\mbg_{uv}$ as
\begin{equation}\label{eq::closeness}
s_r(u, v) \coloneqq 
\begin{cases}
\dfrac{\exp(\bmmu_r\trans \mbg_{uv})}{\sum\limits_{\tilde{v} \in \mc{P}^r_{u \ast}} \exp(\bmmu_r\trans \mbg_{u \tilde{v}}) + \sum\limits_{\tilde{u} \in \mc{P}^r_{\ast v}} \exp(\bmmu_r\trans \mbg_{\tilde{u} {v}})}, & (u, v) \in \mc{P}^r,  \\
0, & (u, v) \notin \mc{P}^r,
\end{cases}
\end{equation}
where $\bmmu_r \in \mathbb{R}^{d_{\mc{V}}}$ is a edge-type--specific vector to be inferred that represents the metric coupled with this type.
Edge types with compatible semantics are expected to share similar $\bmmu_r$, while incompatible edge types make use of different $\bmmu_r$ to avoid the embedding learning on respective semantics to dampen each other.

To measure the capability of the learned embedding in reconstructing the input HIN, the difference between the observed weights and the typed closeness inferred from embedding are used, which leads to the objective to be minimized for edge type $r$
\begin{equation}\label{eq::obj-r}
\mc{O}^r = \mathit{KL}(W^{(r)}_{uv}, s_t(u, v)) = - \sum_{(u, v) \in \mc{P}^r} W^{(r)}_{uv} \log s_r(u, v) + \mathit{const},
\end{equation}
where $\mathit{KL}(\cdot)$ stands for the Kullback-Leibler divergence.

Further, substituting Eq.~\eqref{eq::closeness} into Eq.~\eqref{eq::obj-r} and taking all edge types into account and, the overall objective function becomes
\begin{equation}\label{eq::obj}
\mc{O} =  - \sum_{\substack{(u, v) \in \mc{P}^r \\ r \in \mc{R}}} W^{(r)}_{uv} \log \tfrac{\exp(\bmmu_r\trans \mbg_{uv})}{\sum\limits_{\tilde{v} \in \mc{P}^r_{u \ast}} \exp(\bmmu_r\trans \mbg_{u \tilde{v}}) + \sum\limits_{\tilde{u} \in \mc{P}^r_{\ast v}} \exp(\bmmu_r\trans \mbg_{\tilde{u} {v}})}.
\end{equation}


To formulate edge embeddings required by Eq.~\eqref{eq::obj}, we derive from the same embeddings of the associated nodes regardless of the involved edge type, so that we reach a unified model where the learning process involving multiple edge types can work together and mutually enhance each other if they embody compatible semantics.
While there are many options to build edge embedding from node embedding, we expect our formulation not to be over-complicated, so that the overall model could be computationally efficient and thereby easy-to-use.
Moreover, in order for \heer to handle general HINs, it must also be able to handle directed and undirected edges accordingly.
Considering these requirements, we decompose node embedding into two sections $\mbf_u = \left[ \substack{\mbf_u^O \\ \mbf_u^I} \right]$, where $\mbf_u^O$ and $\mbf_u^I$ are two column vectors of the same dimension, and build edge embedding on top of node embedding as
\begin{equation}\label{eq::edge-rep}
\mbg_{uv} \coloneqq
\begin{cases}
2 \cdot \mbf_u^O \circ \mbf_v^I, & \text{directed representation from } u \text{ to } v,  \\
\mbf_u^O \circ \mbf_v^O + \mbf_u^I \circ \mbf_v^I, & \text{undirected representation},
\end{cases}
\end{equation}
where $\circ$ represents the Hadamard product.
Besides Hadamard product, on can also build $\mbg_{uv}$ in a way similar to Eq.~\eqref{eq::edge-rep} using addition, subtraction, or outer-product.
We leave the exploration of this direction to future works.

Taking Eq.~\eqref{eq::edge-rep} into account, learning node and edge embedding from an HIN by minimizing Eq.~\eqref{eq::obj} is equivalent to the following optimization problem
\begin{equation}\label{eq::opt}
\min_{\{\mbf_{u}\}_{u \in \mc{V}}, \{\bmmu_r\}_{r \in \mc{R}}} \mc{O}.
\end{equation}

\subsection{Model Inference}
The \heer model in Eq.~\eqref{eq::opt} that we aim to infer can be structured as a neural network as illustrated in Figure~\ref{fig::nn}, where $\bm{F} = \left[\mbf_1, \mbf_2, \ldots, \mbf_{|\mc{V}|}\right] \in \mathbb{R}^{d_{\mc{V}} \times |\mc{V}|}$ and $\bm{M} = \left[\bmmu_1, \bmmu_2, \ldots, \bmmu_{|\mc{R}|}\right] \in \mathbb{R}^{d_{\mc{E}} \times |\mc{R}|}$.
Each pair of nodes gets their respective embeddings through the dense layer $\bm{F}$, which further compose edge embedding by function $g(\cdot, \cdot)$. 
The raw scores for all edge types are obtained through another dense layer $\bm{M}$, followed by a type filter where the neuron for an edge type is connected to its corresponding neuron in the next layer only if this type is compatible with the node types of the input node pairs. 
Lastly, the loss is calculated by the typed closeness and the existence of edges in between the input node pair.

Since it is computationally expensive to compute the denominator in Eq.~\eqref{eq::closeness}, we adopt the widely used negative sampling method~\cite{mikolov2013distributed}, which enjoys linear-time computation.
Specifically, each time, an edge between $(u, v)$ with type $r$ is sampled from the HIN with probability proportional to its weight. 
Then $K$ negative node pairs $(u, \tilde{v}_i)$ and $K$ negative node pairs $(\tilde{u}_i, v)$ are sampled, where each $\tilde{u}_i$ has the same type as $u$ and each $\tilde{v}_i$ has the same type as $v$.
The loss function computed from this sample becomes
$$\log \sigma(\bmmu_r\trans \mbg_{uv}) + \sum_{i=1}^K \mathbb{E}_{\tilde{v}_i} \log \sigma(- \bmmu_r\trans \mbg_{u \tilde{v}_i}) + \sum_{i=1}^K \mathbb{E}_{\tilde{u}_i} \log \sigma(- \bmmu_r\trans \mbg_{\tilde{u}_i v}), $$
where $\sigma(\cdot)$ is the sigmoid function $\sigma(x) = \exp(x)/\big(1+\exp(x)\big)$. 

We adopt mini-batch gradient descent with the PyTorch implementation to minimize the loss function with negative sampling, where each mini-batch contains $B$ sampled edges.
We also use the node embeddings pretrained by the homogeneous network embedding algorithm LINE~\cite{tang2015line} to initialize the node embeddings in \heer.
The edge-type--specific scoring vector $\bmmu_r$ is initialized to be all-one vectors. 


\section{Experiments}\label{sec::exp}
In this section, we evaluate the embedding quality of the proposed method and analyze the utility of employing edge representation and heterogeneous metric using two large real-world HINs. 
We first perform an edge reconstruction task to directly quantify how well the embedding algorithms can preserve the information in the input HINs.
Then, we conduct in-depth case studies to analyze the characteristics of the proposed method.

\subsection{Baselines}
We compare the proposed \heer algorithm with baseline methods that fit the setting of our problem, \ie, the methods should be applicable to general HINs without the help of additional supervision or expertise.
\begin{itemize}
\item
\textbf{Pretrained} (LINE~\cite{tang2015line}). 
This baseline uses the LINE algorithm to generate node embeddings, which are also used to initialize \heer.
LINE is a homogeneous network embedding algorithm based on the skip-gram model~\cite{mikolov2013distributed}.
We use inner product to compute the score of observing an edge between a pair of node embeddings following the original paper~\cite{tang2015line}.
\item
\textbf{AspEm}~\cite{shi2018aspem}.
AspEm is a heterogeneous network embedding method that captures the incompatibility in HINs by decomposing the input HIN into multiple aspects with an unsupervised measure using dataset-wide statistics.
Embeddings are further learned independently for each aspect.
This method considers the incompatibility in HINs but does not model different extent of incompatibility.
Furthermore, it does not allow joint learning of embeddings across different aspects.
Out of fairness, we let the number of aspects in AspEm to be two, in order to generate the final embedding with dimension that is identical to other methods.
Inner product is also used to compute the score for this baseline.
\item
\textbf{UniMetrics} (metapath2vec++~\cite{dong2017metapath2vec}). 
This is a partial model of \heer, where the metrics $\{\bm{\mu}_r\}_{r \in \mc{R}}$ are not updated in the training process, \ie, they remain uniform as initialized.
It is equivalent to the metapath2vec++~\cite{dong2017metapath2vec} using all edges as length-1 meta-paths without further selection.
This method restricts the negative sampling to be done within the consistent node types, \ie, performs heterogeneous negative sampling, but still embeds all nodes into the same metric space regardless of types.
\item
\textbf{Pretrained $+$ Logit}. 
On top of the embeddings from the previous Pretrained model, we train a logistic regression (Logit) model for each edge type using the input network.
Then, we compute scores for test instances of each edge type using the corresponding Logit model.
This method models heterogeneous metrics but does not allow the node embeddings and the edge embeddings to be further improved according to the inferred metrics.
\end{itemize}

\subsection{Data Description and Experiment Setups}
In this section, we describe the two real-world HINs used in our experiments as well as experiment setups.

\begin{figure}[t]
 \centering\includegraphics[width=0.9\linewidth]{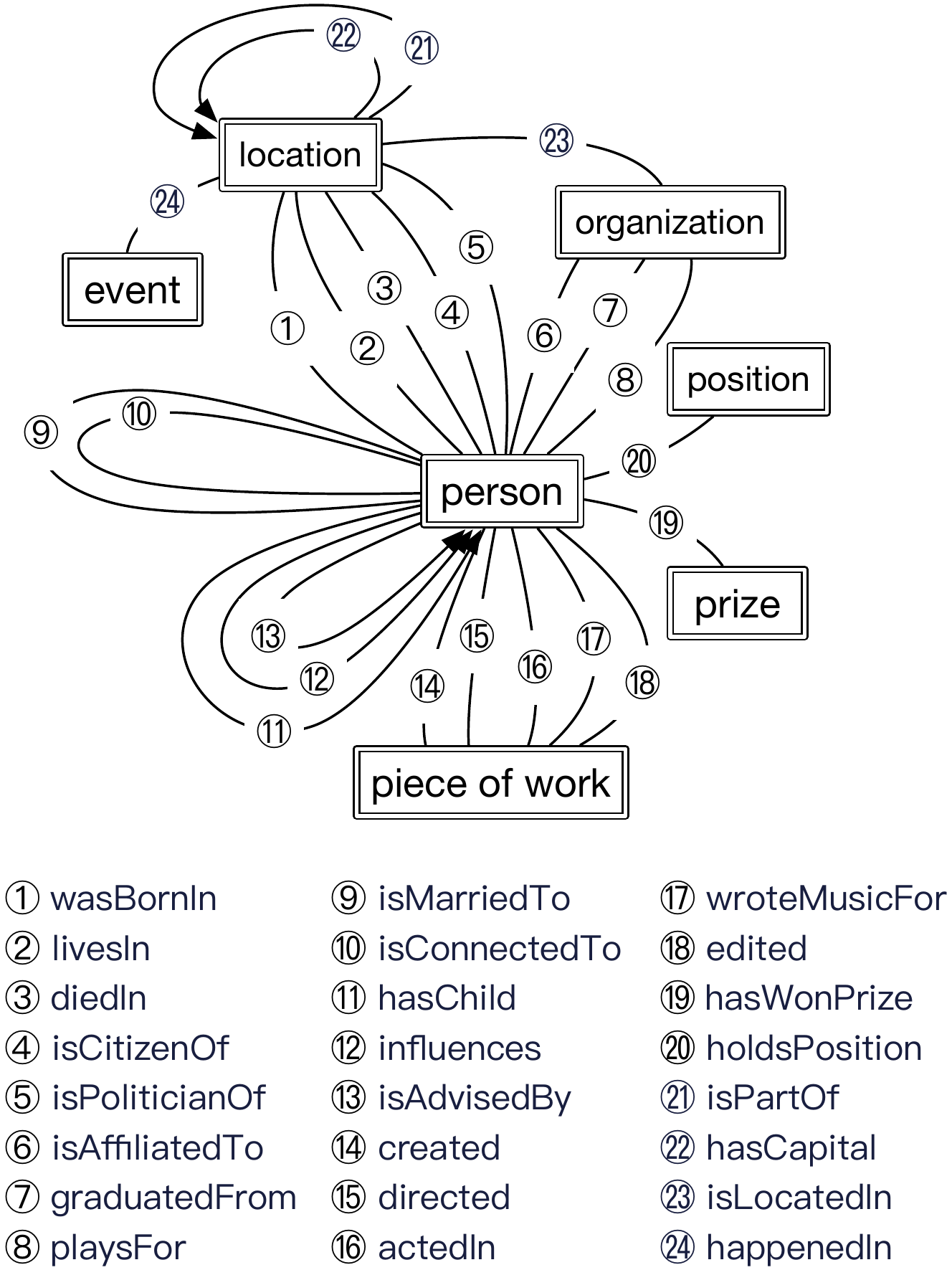}
 \caption[]{The schema of the YAGO network.}\label{fig::yago-schema}
\end{figure}

\vpara{Datasets.}
We use two publicly available real-world HIN datasets: DBLP and YAGO.
\begin{itemize}
\item
\textbf{DBLP} is a bibliographical network in the computer science domain~\cite{tang2008arnetminer}. 
There are five types of nodes in the network: author, paper, key term, venue, and year. 
The key terms are extracted and released by Chen et al.~\cite{chen2017task}.
The edge types include authorship (aut.), term usage (term), publishing venue (ven.), and publishing year (year) of a paper, and the reference relationship from a paper to another (ref.).
We consider the first four edge types as undirected, and the last one as directed.
The corresponding network schema is depicted in Figure~\ref{fig::dblp-schema} on page \pageref{fig::dblp-schema}.
\item
\textbf{YAGO} is a large-scale knowledge graph derived from Wikipedia, WordNet, and GeoNames~\cite{suchanek2007yago}.
There are seven types of nodes in the network: person, location, organization, piece of work, prize, position, and event.
A total of $24$ edge types exist in the network, with five being directed and others being undirected.
These edge types are illustrated together with the schema of the network in Figure~\ref{fig::yago-schema}.
\end{itemize}
We summarize the statistics of the datasets including the total number of nodes, the total number of edges, and the counts of each node type in Table~\ref{tab::data-stat}.

\begin{table}[t]
\centering
    \caption{Basic statistics for the DBLP and YAGO networks.}\label{tab::data-stat}
\scalebox{.9}{
\begin{tabular}{| c || c | c || c | c |}
\hline
Dataset & Node & Edge & Node type & Edge type   \\ \hline
DBLP   & 3,170,793 & 27,126,718 & 5 & 5    \\ \hline
YAGO  & 579,721 & 2,191,464 & 7 & 24\\  \hline
\end{tabular}
}
\end{table}

\begin{table*}[t]
\centering
\caption{Per-edge-type, micro-average, and macro-average MRR achieved by each model in the edge reconstruction task.}\label{tab::edge-rec}
\resizebox{\textwidth}{!}{
\begin{tabular}{l | c | c | c | c | c | c | c | c | c}
\toprule \hline
Dataset &  \multicolumn{7}{c|}{DBLP}  &  \multicolumn{2}{c}{YAGO} \\ \hline
Metric (MRR) & Aut. & Term & Ref. & Pub. venue   & Pub. year & Micro-avg. & Macro-avg. &    Micro-avg. &  Macro-avg.  \\ \hline \hline
Pretrained (LINE~\cite{tang2015line}) & 0.7053 & 0.4830 & 0.8729 & 0.7488 & 0.4986 & 0.6307 & 0.6617  & 0.7454 & 0.6890 \\ \hline
AspEm~\cite{shi2018aspem} & 0.7068 & 0.6010 & 0.8648 & 0.7612 & 0.6791 & 0.6976 & 0.7225 & 0.7832 & 0.6825   \\ \hline
UniMetrics (len-1 metapath2vec++~\cite{dong2017metapath2vec}) & 0.7040 & 0.5772 & 0.8466 & 0.7534 & 0.6781 & 0.6812 & 0.7119 & 0.7437 & 0.6884  \\ \hline 
Pretrained $+$ Logit & 0.8187 & 0.6996 & 0.8072 & 0.8379 & 0.4889 & 0.7310 & 0.7304  & 0.8233 & 0.7012 \\ \hline \hline
\heer & \textbf{0.8964} & \textbf{0.7188} & \textbf{0.9573} & \textbf{0.9132} & \textbf{0.7421}& \textbf{0.8189} & \textbf{0.8456}  & \textbf{0.8635} & \textbf{0.7185}  \\ \hline
\bottomrule
\end{tabular}
}
\end{table*}

\vpara{Experiment Setups.}
For all experiments and all methods, we set the total embedding dimension to be $256$.
That is, for \heer and its related baselines, $\mbf_u \in \mathbb{R}^{256}$ and $\mbf_u^I, \mbf_u^O \in \mathbb{R}^{128}$, and each of the two aspects in AspEm uses a $128$-dim embedding space.
The pretrained model is always tuned to the best according to the performance in the edge reconstruction task to be introduced in Section~\ref{sec::edge-rec}.
The negative sampling rate is always set to $K = 5$ for all applicable models.
We always rescale the pretrained embedding by a constant factor of $0.1$ before feeding them into \heer to improve the learning of heterogeneous metrics, which shares intuition with a previous study~\cite{nickel2017poincare} in improving angular layout at the early stage of model training.
The learning rate for gradient descent for \heer is set to $10$ on both datasets.
Note that we use the same set of hyperparameters for \heer on both DBLP and YAGO in order to provide an easy-to-use solution to the problem of comprehensive transcription of HINs without the hassle of extensive parameter tuning.

\subsection{Edge Reconstruction Experiment}\label{sec::edge-rec}
In order to directly quantify the extent to which an embedding algorithm can preserve the information in the input HINs, we devise the edge reconstruction experiments for both datasets.
For each HIN, we first knock out a portion of edges uniformly at random, with a certain knock-out rate $\kappa \in (0, 1)$.
Embedding of the network after knock-out is then learned using each compared method.
The task is to reconstruct the edges being knocked out using the learned embedding models.

Specifically, for each edge that is knocked out from the network, suppose it is between node pair $(u, v)$ and of edge type $r$, we randomly sample $10$ negative pairs $(u, \tilde{v})$ that do not have type-$r$ edges in the original full network, where $\tilde{v}$ is of the same node type as $v$.
For any model after training, a score can be calculated to reflect the likelihood for each of the $11$ node pairs to be associated by type-$r$ edge in the current model.
The reciprocal rank is then computed to measure the quality of the model, where the reciprocal rank is the reciprocal of the rank of the positive pair among all $11$ node.
Similarly, another reciprocal rank is computed for the same node pair $(u, v)$ and $10$ other randomly sampled negative pairs $(\tilde{u}, v)$ with fixed $v$ but sampled $\tilde{u}$.
Finally, we report the mean reciprocal rank (MRR), which is computed by the mean of reciprocal ranks for the target test instances.
In particular, the micro-average MRR and the macro-average MRR are reported for both DBLP and YAGO, where the micro-average MRR is computed by the mean of all reciprocal ranks computed regardless of edge types, while the macro-average MRR is derived by first computing the mean of reciprocal ranks for each edge type, and then averaging all these means across different edge types.
Additionally, we also report the MRR for each edge type for DBLP, since DBLP involves only $5$ edge types, while YAGO has as many as $24$ edge types.
We present the results with knock-out rate $\kappa = 0.4$ in Table~\ref{tab::edge-rec}.

\vpara{Modeling incompatibility benefits embedding quality.}
As shown in Table~\ref{tab::edge-rec}, the proposed \heer model outperformed all baselines in both datasets under both micro-average MRR and macro-average MRR, which demonstrated the effectiveness of the proposed method.
Even when looking at each edge type in DBLP, the MRR achieved by \heer was still the best.
Besides, in DBLP, AspEm outperformed Pretrained and UniMetrics on most metrics.
Recall that AspEm decomposed the HIN into distinct aspects using dataset-wide statistics.
As a result, it forbade semantically incompatible edge types to negatively affect each other in the embedding learning process and thereby achieved better results.
In YAGO, the baselines considering heterogeneity did not always clearly outperform the simplest baseline Pretrained.
We interpret this result by that YAGO has much more edge types than DBLP, which introduces even more varied extent of incompatibility, and the relatively simple approaches adopted by AspEm and UniMetrics in modeling incompatibility may not be enough to bring in significant performance boost.
In contrast, armed with heterogeneous metrics fine-grained to the edge type level, \heer outperformed Pretrained by a clear margin even in YAGO.

\vpara{Heterogeneous metrics helps improving embedding quality.}
As a sanity check, the Pretrained $+$ Logit model helps rule out the possibility that \heer archives better results only by learning edge-type--specific metrics without actually improving embedding quality.
From Table~\ref{tab::edge-rec}, it can be observed that by coupling with the additional edge-type--specific logistic regression and modeling heterogeneous metrics, the performance was improved on top of the Pretrained mode.
This observation further consolidated the necessity of employing heterogeneous metrics for different edge types in solving the problem of comprehensive transcription of heterogeneous information networks.
However, Pretrained $+$ Logit still performed worse than the proposed \heer mode, which implies that the inferred heterogeneous metrics of \heer indeed in return improved the quality of the node and edge embedding.

\begin{figure}[t]
  \centering
  \begin{subfigure}[t]{\linewidth}
    \centering\includegraphics[width=\linewidth]{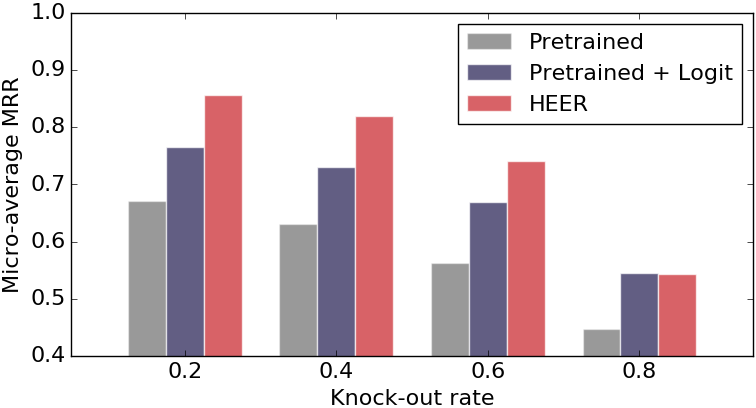}
    \caption{DBLP.}
  \end{subfigure}
  \begin{subfigure}[t]{\linewidth}
    \centering\includegraphics[width=\linewidth]{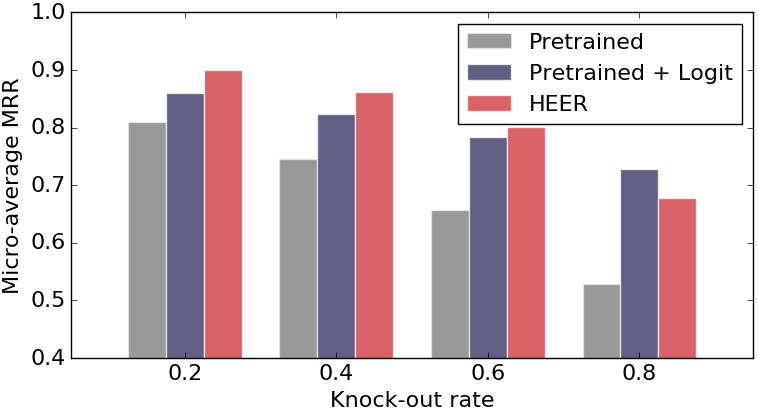}
    \caption{YAGO.}
  \end{subfigure}
  \caption{Micro-average MRR under multiple knock-out rate $\kappa$ in the egde reconstruction tasks.}\label{fig::varying-ko-rate}
\end{figure}

\begin{figure*}[t]
 \centering\includegraphics[width=\linewidth]{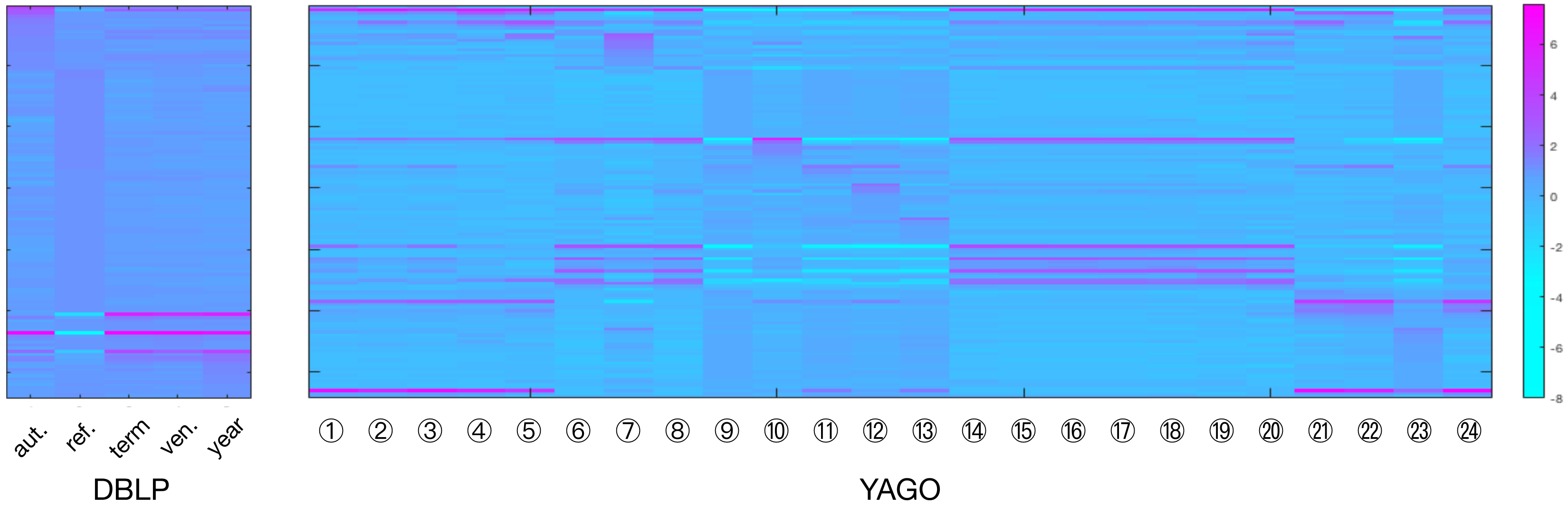}
 \caption[]{The learned heterogeneous metrics of the \heer model sensed the heterogeneous semantics of the HINs.}\label{fig::heatmap}
\end{figure*}

\vpara{Heterogeneous negative sampling is not always enough to capture incompatibility.}
UniMetric performed better than the Pretained model in DBLP, it failed to make an absolute win over Pretained as other methods did in the YAGO dataset.
Our interpretation of this result is that while the heterogeneous negative sampling as used by UniMetric does leverage certain type-specific information in HINs, it may not be always enough to model incompatibility and resolve the negative impact it brings to embedding quality.
This observation is to be further corroborated in Section~\ref{sec::case-studies} by examining the capability of transcribing information implied by meta-paths in each model.

\vpara{Varying Knock-Out Rate in Edge Reconstruction}.
Additionally, we vary the knock-out rate $\kappa$ on both datasets and report the micro-average MRR for the proposed \heer model and two baseline models that require less training time.
As presented in Figure~\ref{fig::varying-ko-rate}, \heer outperformed all baselines under at most knock-out rates, which demonstrated the robustness of the proposed model.
Besides, Pretrained $+$ Logit outperformed Pretrained at all knock-out rates, which is also in line with the previous results we have presented.
Notably, \heer did not outperform Pretrained $+$ Logit when the knock-out rate $\kappa = 0.8$.
This is explainable because only a very small portion ($20\%$) of the original HIN was used for learning embedding when $\kappa = 0.8$. 
With a bigger model size than Pretrained $+$ Logit, \heer was more prone to suffering from over-fitting.

\subsection{Case Studies}\label{sec::case-studies}
In this section, we conduct a series of in-depth case studies to understand the characteristics of the proposed \heer model.

\vpara{Learned heterogeneous metrics.} 
\heer leverages heterogeneous metrics to model the different extent of incompatible semantics carried by different edge types.
In this section, we analyzed the learned metrics $\{\bm{\mu}_r\}_{r \in \mc{R}}$ in \heer to verify if they indeed captured the different semantics and thereby enriched the model capability.

To this end, we use heat maps to illustrate $\{\bm{\mu}_r\}_{r \in \mc{R}}$ that are learned in the edge reconstruction experiments on both HINs.
Specifically, for each dataset, we first standardize the elements of each $\bm{\mu}_r$ to have zero mean and unit deviation, so that $\bm{\mu}_r$'s have comparable scales after standardization for all $r \in \mc{R}$.
Then, we re-order the $d_{\mc{E}}$ dimensions for better visualization and plot the heat maps in Figure~\ref{fig::heatmap}.

{
\newcommand*{\circledsix}{\tikz[baseline=(char.base)]{\node[shape=circle,draw,inner sep=1.2pt] (char) {6};}}
\newcommand*{\circledeight}{\tikz[baseline=(char.base)]{\node[shape=circle,draw,inner sep=1.2pt] (char) {8};}}
\newcommand*{\circlednine}{\tikz[baseline=(char.base)]{\node[shape=circle,draw,inner sep=1.2pt] (char) {9};}}
\newcommand*{\circledeleven}{\tikz[baseline=(char.base)]{\node[shape=circle,draw,inner sep=.2pt] (char) {11};}}
\newcommand*{\circledtwelve}{\tikz[baseline=(char.base)]{\node[shape=circle,draw,inner sep=.2pt] (char) {12};}}
Recall that the inferred metrics $\{\bm{\mu}_r\}_{r \in \mc{R}}$ were set to be all-one vectors in initialization, whereas Figure~\ref{fig::heatmap} shows that different metrics have generally reached different distributions over the $d_{\mc{E}}$ dimensions after training.
This implies that the inferred metrics of the \heer model indeed sensed the heterogeneity of the HINs, and were using different projected metric spaces to embed different semantics of the input network.
Notably, it can be seen from the heat map of YAGO that edge types \circledsix{} (isAffiliatedTo) and \circledeight{} (playsFor) have similar inferred metrics.
This is actually expected because these edge types are often associated with the relationship between professional sports players and their associated teams in YAGO.
Besides, similar phenomenon can be observed between \circlednine{} (isMarriedTo) and edge type \circledeleven{} (hasChild).

\begin{figure}[t]
  \centering\includegraphics[width=\linewidth]{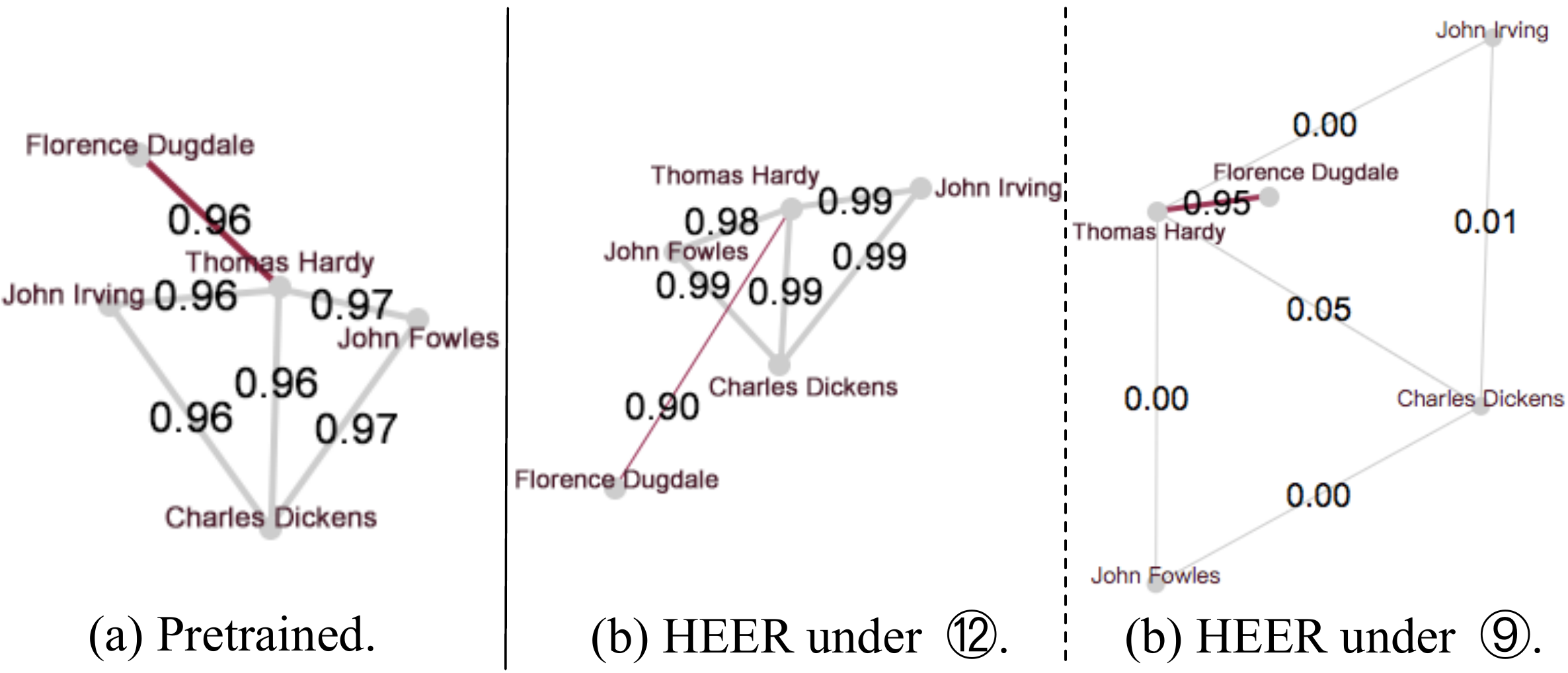}
  \caption{The subnetwork surrounding Thomas Hardy in YAGO in multiple embedding models under potentially different metrics.}\label{fig::dickens}
\end{figure}

\vpara{Embedded subnetwork with different edge types.}
In order to understand how the network is impacted by the introduction of heterogeneous metrics, we took a closer look at a subnetwork surrounding the British writer Thomas Hardy in the YAGO dataset. 
Multiple other writers having the \textit{influences} relationship (\circledtwelve{}, colored gray) with Thomas Hardy include Charles Dickens, John Fowles, and John Irving, while Florence Dugdale and Thomas Hardy enjoyed the \textit{isMarriedTo} relationship (\circlednine{}, colored red).
Besides, Fowles and Irving are also influenced by Dickens.
In Figure~\ref{fig::dickens}, we visualized this subnetwork under each embedding model with the inferred possibility of edge existence marked, where the embedding models are training using the entire network.
It can be seen from Figure~\ref{fig::dickens}a that without distinguishing edge types, Pretrained assigned high possibilities to all edges.
Meanwhile, with the learned heterogeneous metrics, the \heer model assigned a relatively low probability for Dugdale and Hardy under the metric for \textit{influences} as in Figure~\ref{fig::dickens}b (note that Dugdale was also a writer), and a clearly higher probability under the metric for \textit{isMarriedTo} as in Figure~\ref{fig::dickens}c.
}

\begin{figure}[t]
  \centering
  \begin{subfigure}[t]{0.46\linewidth}
    \centering\includegraphics[width=\linewidth]{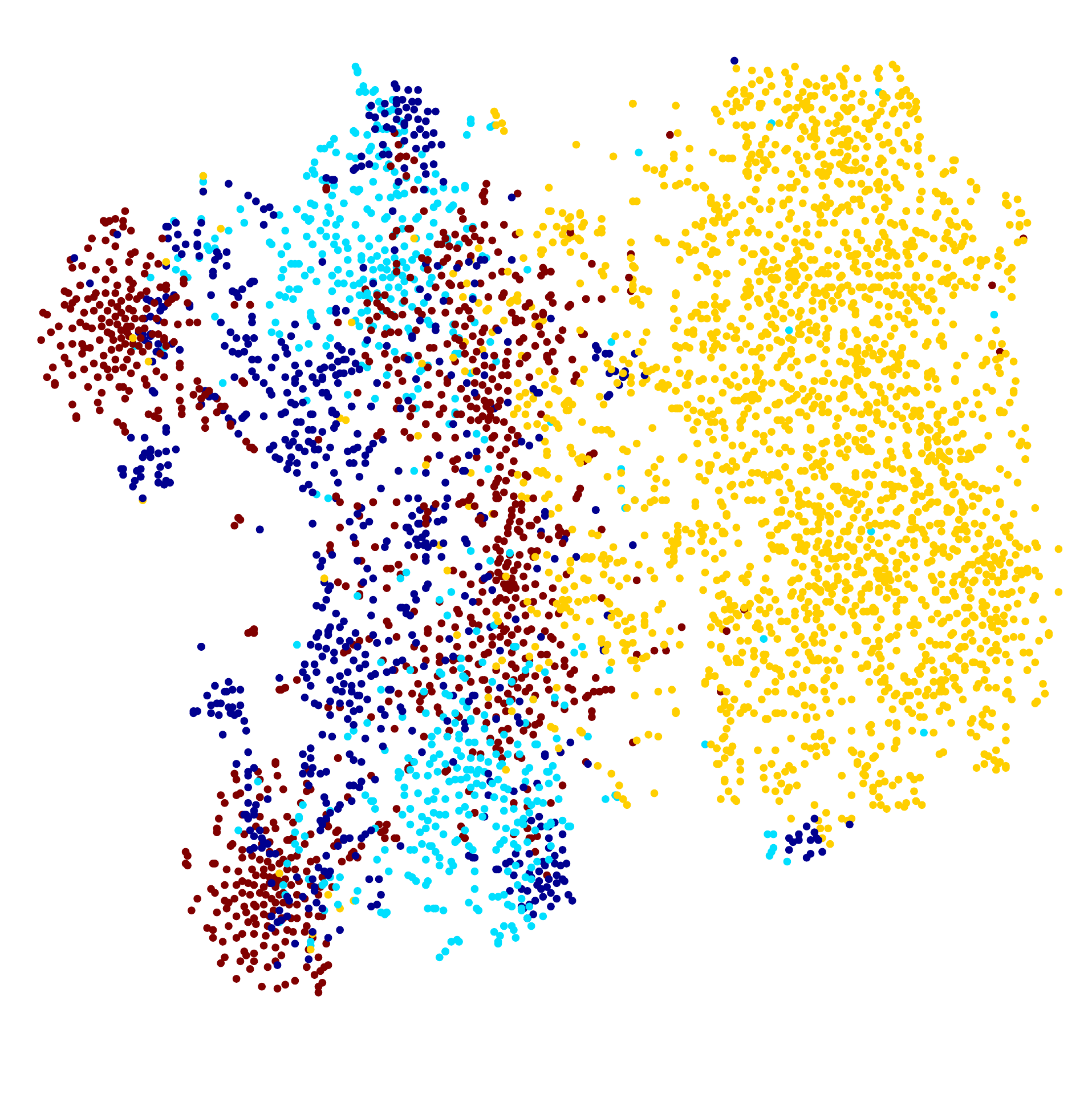}
    \caption{UniMetrics.}\label{fig::tsne-mp2v}
  \end{subfigure}
  \hspace{6pt}
  \begin{subfigure}[t]{0.46\linewidth}
    \centering\includegraphics[width=\linewidth]{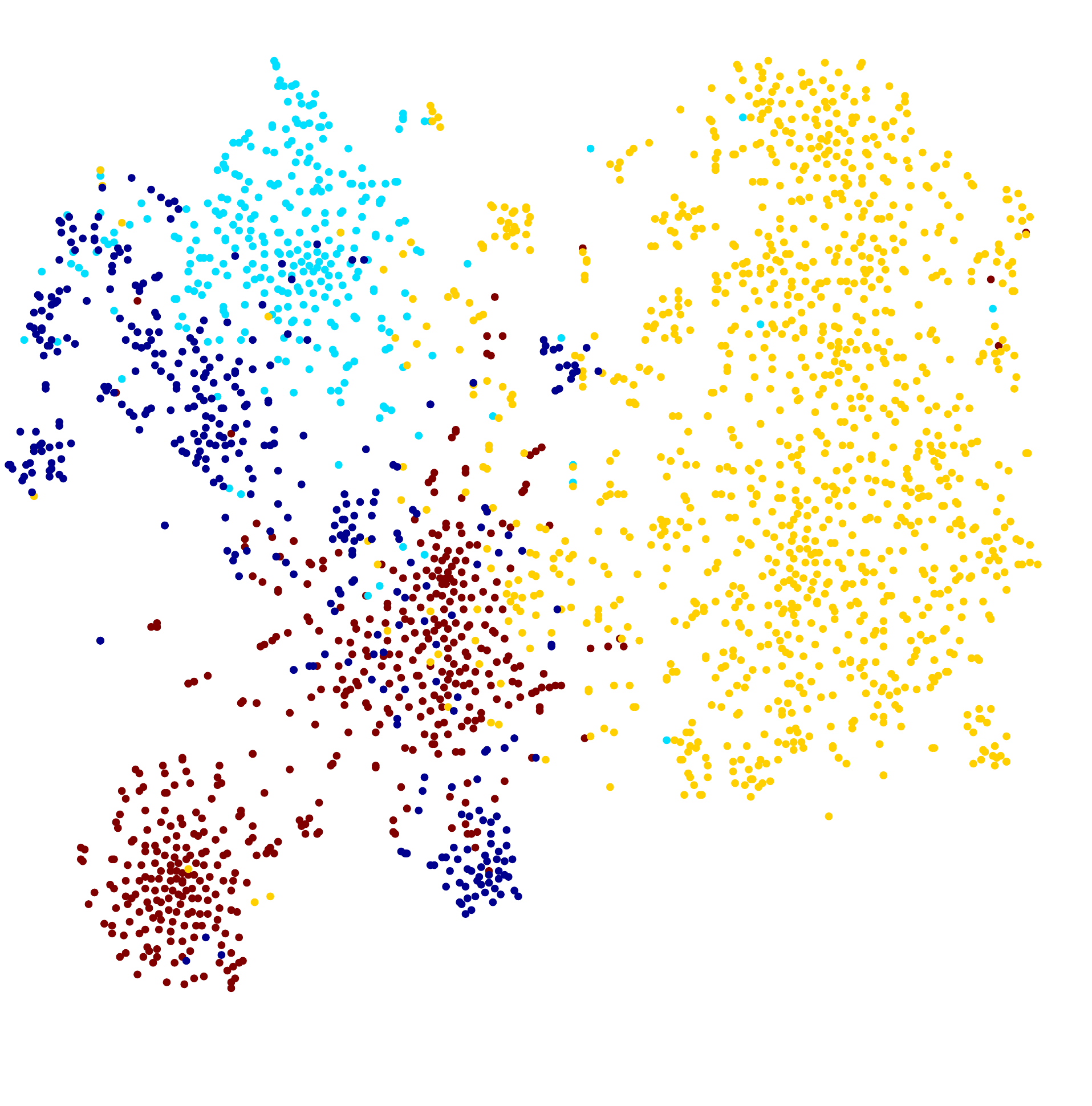}
    \caption{\heer.}\label{fig::tsne-heer}
  \end{subfigure}
  \caption{Visualization using t-SNE of $2,370$ paper nodes that are linked to a given paper via four different meta-paths in DBLP. Each node is colored according to the meta-path it uses to reach the given paper. In Figure~\ref{fig::tsne-heer}, the nodes with same meta-path are clustered together, which implies \heer can preserve the information carried by different meta-paths in embedding learning even without the use of any meta-path. As a comparison, UniMetrics (len-1 metapath2vec++) yields less distinct clusters, with red, cyan, and blue nodes mingled together, showing adopting heterogeneous negative sampling without learned heterogeneous metrics is not sufficient to preserve the heterogeneity in the HIN embedding process.}\label{fig::tsne}
\end{figure}

\vpara{Transcription of information implied by meta-paths.}
When embedding each of the HINs, no meta-paths of length greater than $1$ (\ie., besides edges) were used, where a meta-path is the concatenation of certain edge types used to reflect certain semantics of an HIN~\cite{sun2011pathsim, sun2013mining, shi2017survey}.
We would like to verify whether the input HIN can be transcribed so comprehensively that the signals implied by meta-paths are also preserved in the embedding process, even without the use of guidance from meta-paths.

To this end, we visualize the embedding results considering four different meta-paths: $[\mathrm{paper}]  \dfrac{\text{\footnotesize{ven.}}}{} [\mathrm{venue}] \dfrac{\text{\footnotesize{ven.}}}{} [\mathrm{paper}]$ (colored red), $[\mathrm{paper}] \xrightarrow{\mathrm{ref.}} [\mathrm{paper}]$ (colored cyan), $[\mathrm{paper}] \dfrac{\text{\footnotesize{aut.}}}{}  [\mathrm{author}] \dfrac{\text{\footnotesize{aut.}}}{} [\mathrm{paper}]$ (colored blue), and $[\mathrm{paper}] \dfrac{\text{\footnotesize{year}}}{}  [\mathrm{year}] \dfrac{\text{\footnotesize{year}}}{} [\mathrm{paper}]$ (colored yellow).
In particular, we consider a given paper node and find all paper nodes connected to this given paper by the aforementioned four meta-paths.
Then we visualize the embeddings of all the nodes found in this way using the popular t-Distributed Stochastic Neighbor Embedding (t-SNE)~\cite{maaten2008visualizing} algorithm.
Each node is colored according to the meta-path it uses to connect to the given paper node.
Additionally, we randomly downsampled the group of nodes reached by meta-path $[\mathrm{paper}] \dfrac{\text{\footnotesize{year}}}{}  [\mathrm{year}] \dfrac{\text{\footnotesize{year}}}{} [\mathrm{paper}]$ because a year can have edges with tens of thousands of paper nodes.
The visualization results are shown in Figure~\ref{fig::tsne}.

In Figure~\ref{fig::tsne-heer}, the nodes with the same color are generally clustered together, which implies \heer can preserve the information implied by different meta-paths in embedding learning even without the use of any meta-path. 
As a comparison, we also visualized the same set of nodes in Figure~\ref{fig::tsne-mp2v} using embedding generated by UniMetrics (len-1 metapath2vec++), which yields less distinct clusters, with red, cyan, and blue nodes mingled together.
Recall that UniMetrics also considers edge types when conducting negative sampling, and is different from \heer only in that the former does not employ heterogeneous metrics leaning.
This result again demonstrated that adopting heterogeneous negative sampling without learned heterogeneous metrics is not sufficient to preserve the heterogeneity in the HIN embedding process, and is therefore not ideal for solving the problem of comprehensive transcription of heterogeneous information networks.

\begin{figure}[t]
\centering
        \includegraphics[width=.45\textwidth]{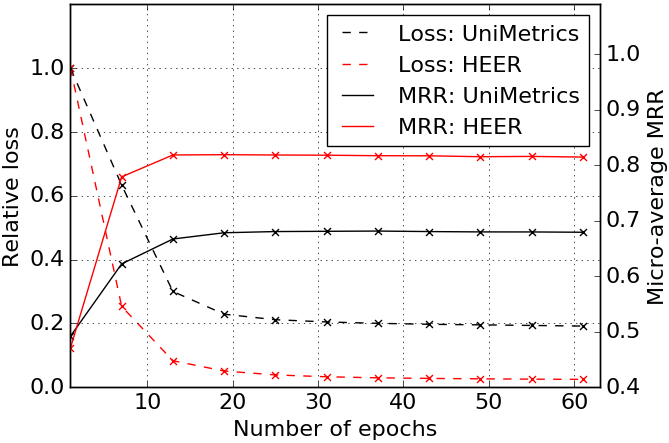}
\caption{The relative loss and the micro-average MRR against the number of epochs for the proposed \heer model and the UniMetrics (len-1 metapath2vec) model.}\label{fig::efficiency}
\end{figure}

\subsection{Efficiency Study}
For efficiency study, we plot out the loss and the performance of the proposed \heer algorithm against the number of epochs in the edge reconstruction experiment.
We also illustrate the same curves of the UniMetrics (len-1 metapath2vec++) model for comparison.
The results are presented in Figure~\ref{fig::efficiency}.

Judging from the curve for the loss again the number of epochs in Figure~\ref{fig::efficiency},  \heer converges at a comparable rate with the skip-gram--based UniMetrics (metapath2vec++).
Besides, \heer took less than twice as much time to finish each epoch as metapath2vec++ did. 
This is expected because \heer only additionally requires one-step gradient descent for one $\bm{\mu_r}$ when training on each sampled training example.
As a result, the time complexity of \heer for each epoch differs from that of metapath2vec++ by a small constant factor.
Combining the above two properties, \heer enjoys overall complexity linear to the number of nodes as skip-gram--based algorithms do~\cite{grover2016node2vec, abu2017learning, dong2017metapath2vec}.


\section{Conclusion and future works}
We studied the problem of the comprehensive transcription of HINs in embedding learning, which preserves the rich information in HINs and provides an easy-to-use approach to unleash the power of HINs.
To tackle this problem, we identify that different extents of semantic incompatibility exist in real-world HINs, which pose challenges to the comprehensive transcription of HINs.
To cope with these challenges, we propose an algorithm, \heer, that leverages edge representations and heterogeneous metrics.
Experiments and in-depth case studies with large real-world datasets demonstrate the effectiveness of \heer and the utility of edge representations and heterogeneous metrics.

With the availability of edge representations proposed this paper, future works include exploration of more loss functions over edge representation, such as regression to model edges associated with ratings on HINs that have user-item reviews, or soft-max to model HINs where at most one edge type can exist between a pair of node types.
One may also explore alternate ways to build edge embedding $\mbg_{uv}$ using addition, subtraction, outer-product, or deeper architectures.
We leave the exploration of this direction to future works.
Besides, it is also worthy of studying further boost the performance of \heer by incorporating higher-order structures such as network motifs, while retaining the advantage of \heer for being able to preserve the rich semantics from HINs.

\vpara{Acknowledgments.}
This work was sponsored in part by U.S. Army Research Lab. under Cooperative Agreement No. W911NF-09-2-0053 (NSCTA), DARPA under Agreement No. W911NF-17-C-0099, National Science Foundation IIS 16-18481, IIS 17-04532, and IIS-17-41317, DTRA HDTRA11810026, and grant 1U54GM114838 awarded by NIGMS through funds provided by the trans-NIH Big Data to Knowledge (BD2K) initiative (www.bd2k.nih.gov). Any opinions, findings, and conclusions or recommendations expressed in this document are those of the author(s) and should not be interpreted as the views of any U.S. Government. The U.S. Government is authorized to reproduce and distribute reprints for Government purposes notwithstanding any copyright notation hereon.

\bibliographystyle{ACM-Reference-Format}
\balance
\bibliography{sigproc}

\end{document}